\documentclass{eas}
\usepackage{graphicx}
\usepackage{sidecap}
\begin{document}

\title{The role of ices in star-forming clouds}
\author{S. Hocuk}
\address{Max-Planck-Instit\"{u}t f\"{u}r extraterrestrische Physik, Giessenbachstrasse 1, 85748 Garching, Germany. \email{seyit@mpe.mpg.de}}
\begin{abstract}
Ices play a critical role during the evolution of interstellar clouds. Their presence is ubiquitous in the dense molecular medium and their impact is not only limited to chemistry. Species adsorbed onto dust grains also affect cloud thermodynamics. It all depends on the interstellar conditions, the chemical parameters, and the composition of ice layers. In this work, I study the formation of ices by focusing on the interplay between gas and solid phase to determine their role on cloud evolution and star formation. I show that while the formation of ices greatly impacts the cloud chemistry, their role on the thermodynamics is more conservative, and their influence on star formation is only marginal.
\end{abstract}
\maketitle
\section*{Introduction}
The presence of dust grains in the intersellar medium is on the order of 1\% by mass (Fischer \etal\ \cite{2014Natur.505..186F}). Their impact, however, is much greater and in some cases dominates the physical processes. Dust enhances or depletes gas-phase molecules and influences the thermodynamics through direct (gas-grain collision) or indirect (altering the abundances) processes. In this, the surface composition of dust grains is crucial. When ice layers start to grow on grain surfaces, there are several things that change with surface chemistry. Amongst other things, the adsorption and diffusion energies of the species residing on grain surfaces change, since the substrate to which species bind changes, the opacity of dust changes, which affects dust temperature, and even how dust coagulates will change, which changes the grain size distribution. Because water ice is usually the main constituent of the icy surface, binding energies are generally higher due to the strong hydrogen bonding. How this affects the gas-phase chemistry, what is the composition of the first ice layer, and whether the observed gas-phase abundances can be explained are all relevant questions.

Considering the changes in the chemistry, one can also anticipate an impact on cloud dynamics. The process of freeze-out, for example, becomes an important factor that indirectly affects cloud cooling. CO ro-/vibrational line cooling is a major coolant in molecular clouds at number densities above a few 10$^3$ cm$^{-3}$ and CO can freeze out efficiently at temperatures below 20 K. These are typical star-forming cloud conditions. One can then ask if CO freeze-out has a strong enough impact on the thermodynamics of a cloud to affect its entire evolution, or, are non-thermal desorption processes, such as photodesorption and chemical desorption, significant enough to counter the cooling deprivation caused by CO freeze-out? Will star formation be affected by this? 

I have tried to answer some of these questions in the past two years from a numerical point of view with several publications Hocuk \& Cazaux \cite{hocuk3} and Hocuk \etal (\cite{hocuk1,hocuk2}). Here I give an overview on the role of ices on molecular abundances and on cloud dynamics, aided by some new results.

\section{How the gas grain interplay impacts cloud chemistry}
\begin{figure*}
\centering
\includegraphics[scale=0.31]{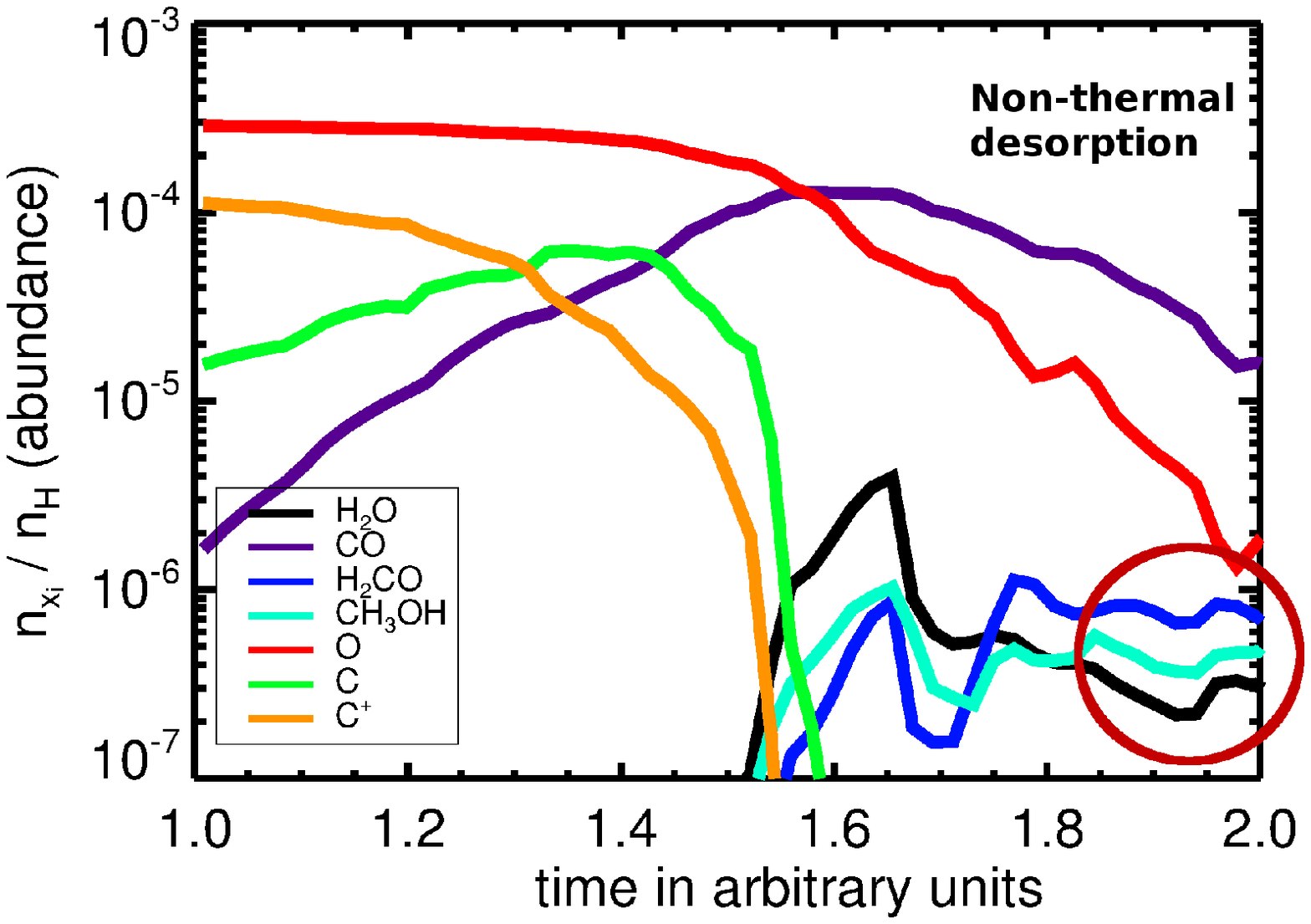}
\includegraphics[scale=0.31]{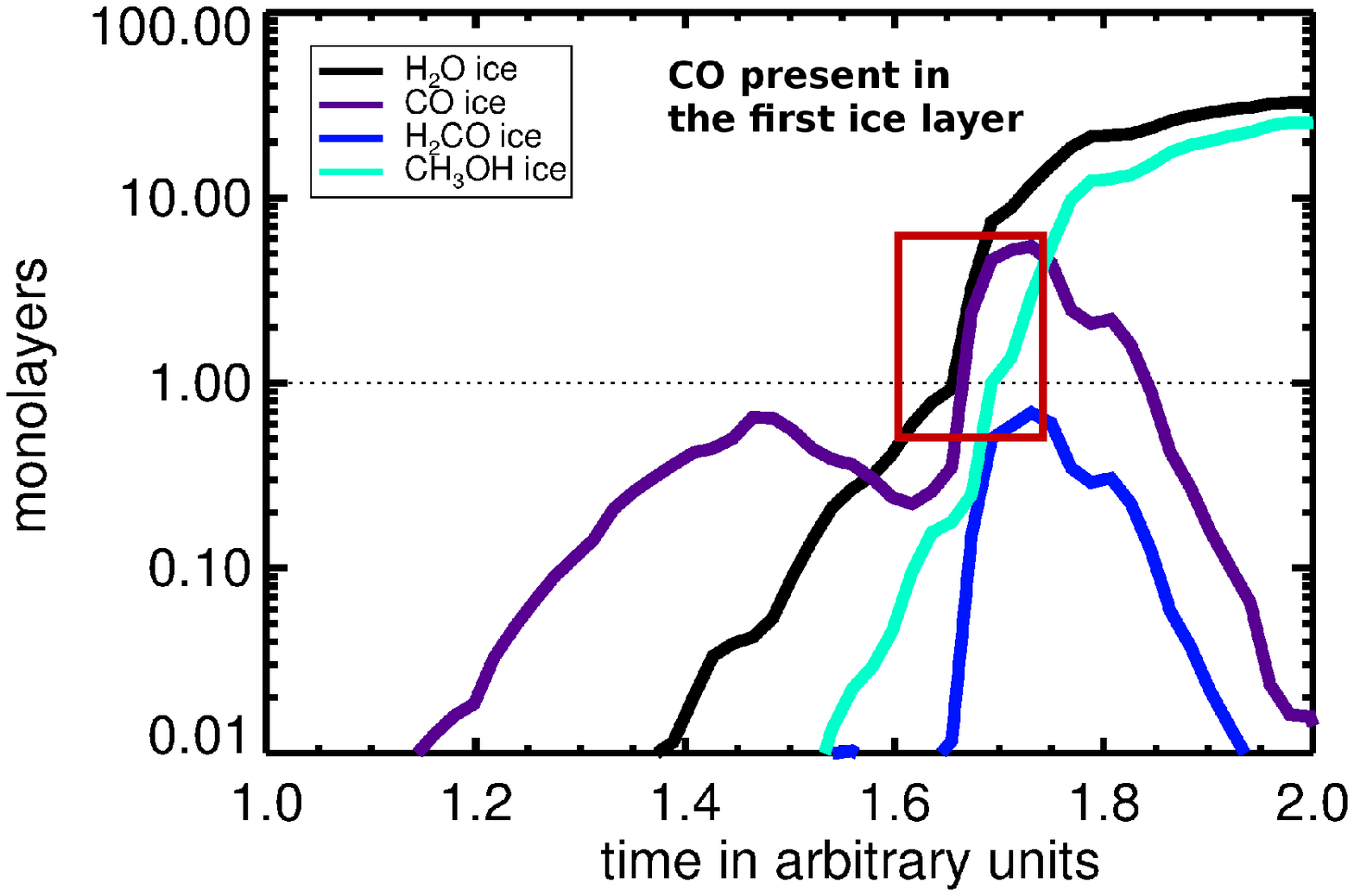}
\caption{Chemical abundances in gas (left) and solid (right) phases as a function of time in a collapsing cloud. Figure adapted from Hocuk \& Cazaux (\cite{hocuk1}).}
\label{fig1}
\end{figure*}

The influence of grain surface chemistry (GSC) on the gas phase and on the formation of the first ice layer is best studied with simulations of diffuse clouds. In the earliest phases of cloud evolution, the composition of the gas will be atomic due to UV background radiation field, which is typically $G_0=1$ in our Galaxy in terms of the Habing field (\cite{habing1968}). In Hocuk \etal\ (\cite{hocuk2}), we performed such a simulation to study the chemical evolution. Figure \ref{fig1} shows the abundances in the two phases, gas and dust.
We can see from this figure that CO is dropping after reaching an abundance of $1.3\times10^{-4}$. This is due to the freeze-out process. We can also see that there is a significant amount of water, formaldehyde, and methanol present in the gas phase after this time. This shows that water forms abundantly on grain surfaces and that CO reacts to form higher order molecules. Here, the key result is that the chemical desorption process is strong enough to release the thermally locked species into the gas phase. From recent higher resolution simulations where I compare models with and without ices in collapsing clouds, I find and confirm that in order to form CO$_2$, H$_2$CO, CH$_3$OH, grain surface chemistry is necessary and that the water and OH abundances are enhanced even when ices are present.

From the surface composition of ices, one can notice that CO, during the formation of the first ice layer, is well mixed with water ice. The high abundance of the CO molecules in the gas phase and the cold dust temperatures make this possible for typical interstellar radiation fields ($G_0 = 1$). The ices start to grow rapidly once one monolayer of ice has formed.

\section{How micro-processes affect cloud dynamics}
\begin{SCfigure}
\centering
\includegraphics[scale=0.31]{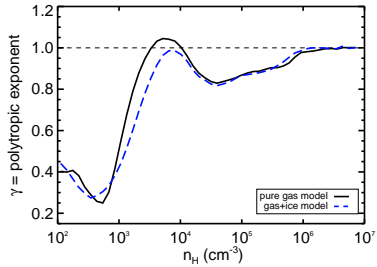}
\caption{The effect of ice formation on the EOS. The biggest impact is that the EOS becomes softer in the density regime $n_{\rm H} = 10^{3}$ - $10^{4.5}$ cm$^{-3}$ due to the ices. The dashed line represents the isothermal condition.}
\label{fig2}
\end{SCfigure}

Any changes in the gas-phase abundances will affect the thermodynamics of a molecular cloud, since the cooling of gas below a number density of $n_{\rm H}\sim10^{4.5}$\,cm$^{-3}$ is dominated by line emission. Between the densities $n_{\rm H}=10^{3.5} - 10^{4.5}$\,cm$^{-3}$, it is CO ro-/vibrational line emission that governs the cooling, however, this molecule is subject to freeze-out in the same density regime. In Hocuk \etal\ (\cite{hocuk1,hocuk3}), we have tested the impact of freeze-out on the thermal balance with simulations, where we saw the consequences on the equation of state (EOS), $P \propto \rho^{\gamma}$. Here, in Figure\,\ref{fig2}, I show the effect that ice formation has on the EOS.

In recent simulations I also find that the filamentary structure may be shaped by the gas-dust collisional coupling. When the dust cooling causes the EOS to reach a minimum at the density $n_{\rm H} \simeq 5\times10^4$\,cm$^{-3}$, this is the moment where the Jeans length reaches the size scale that is typically observed for filaments, $\lambda_{\rm J} \simeq 0.1$\,pc. From this point on, the change in $\gamma$ is positive toward higher densities, thereby suppressing further fragmentation and making this size scale a preferred one. However, since this occurs at $n_{\rm H} \geq 3\times10^4$\,cm$^{-3}$, it is no longer affected by the impact of freeze-out. I find that the impact of ices now come by the change in the opacity of dust, causing the dust temperature to increase by 1 Kelvin at high visual extinctions.

\section{How the chemistry on icy surfaces influences star formation}
\begin{figure*}
\centering
\includegraphics[scale=0.31]{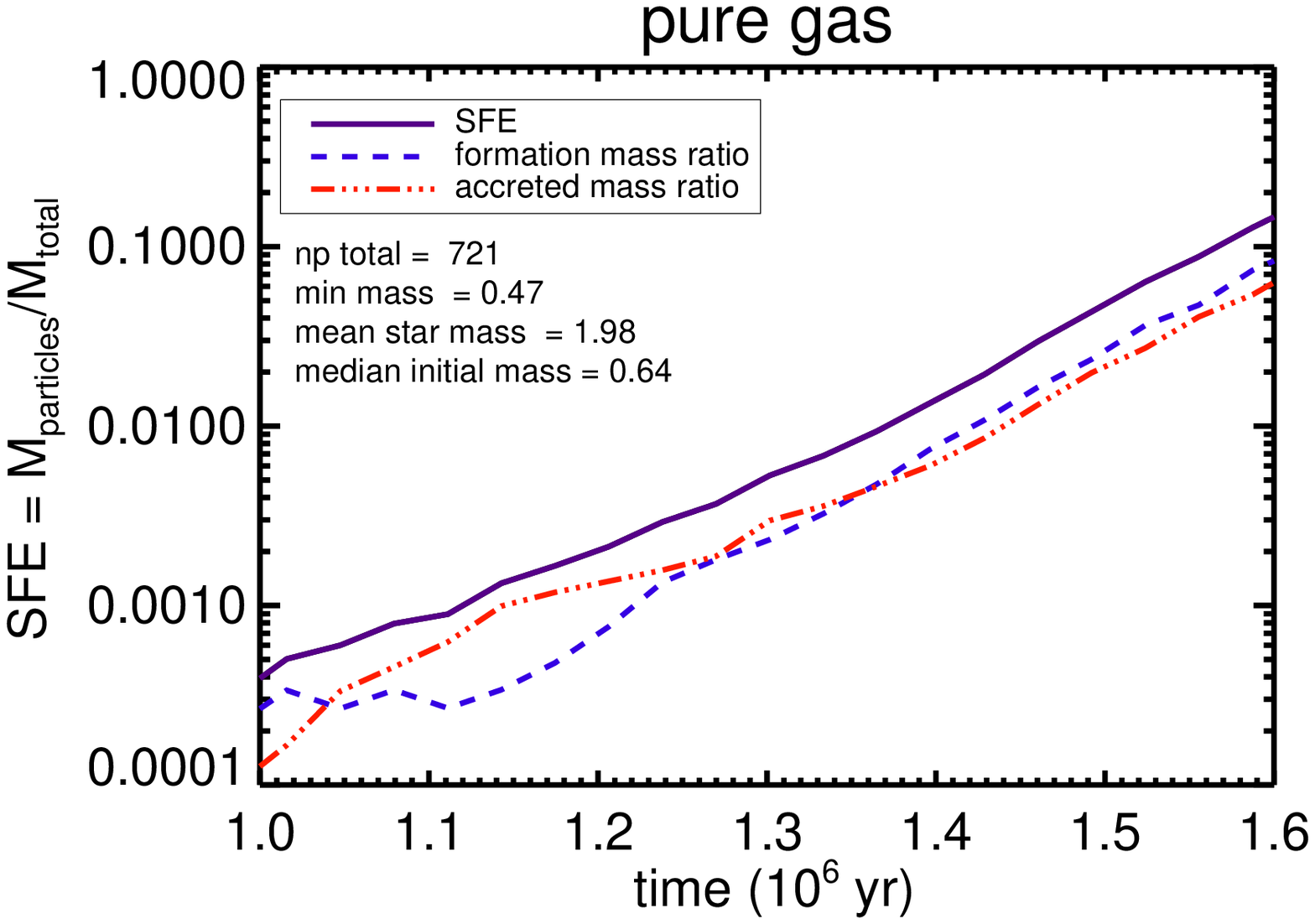}
\includegraphics[scale=0.31]{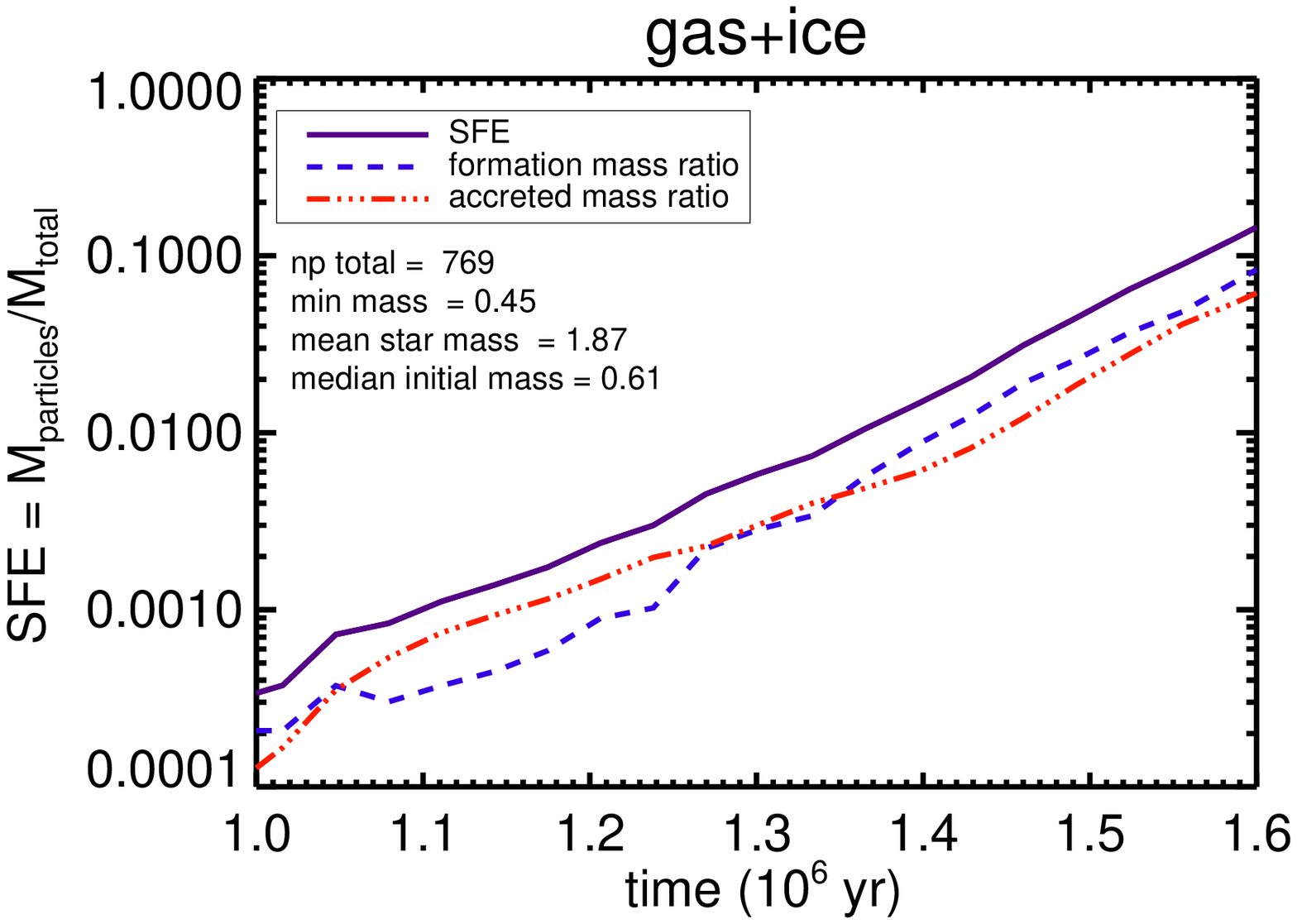}
\caption{The star-formation efficiency (SFE) within a free-fall time ($t_{\rm ff} = 1.63$\,Myr).}
\label{fig3}
\end{figure*}

\begin{figure*}
\centering
\includegraphics[scale=0.31]{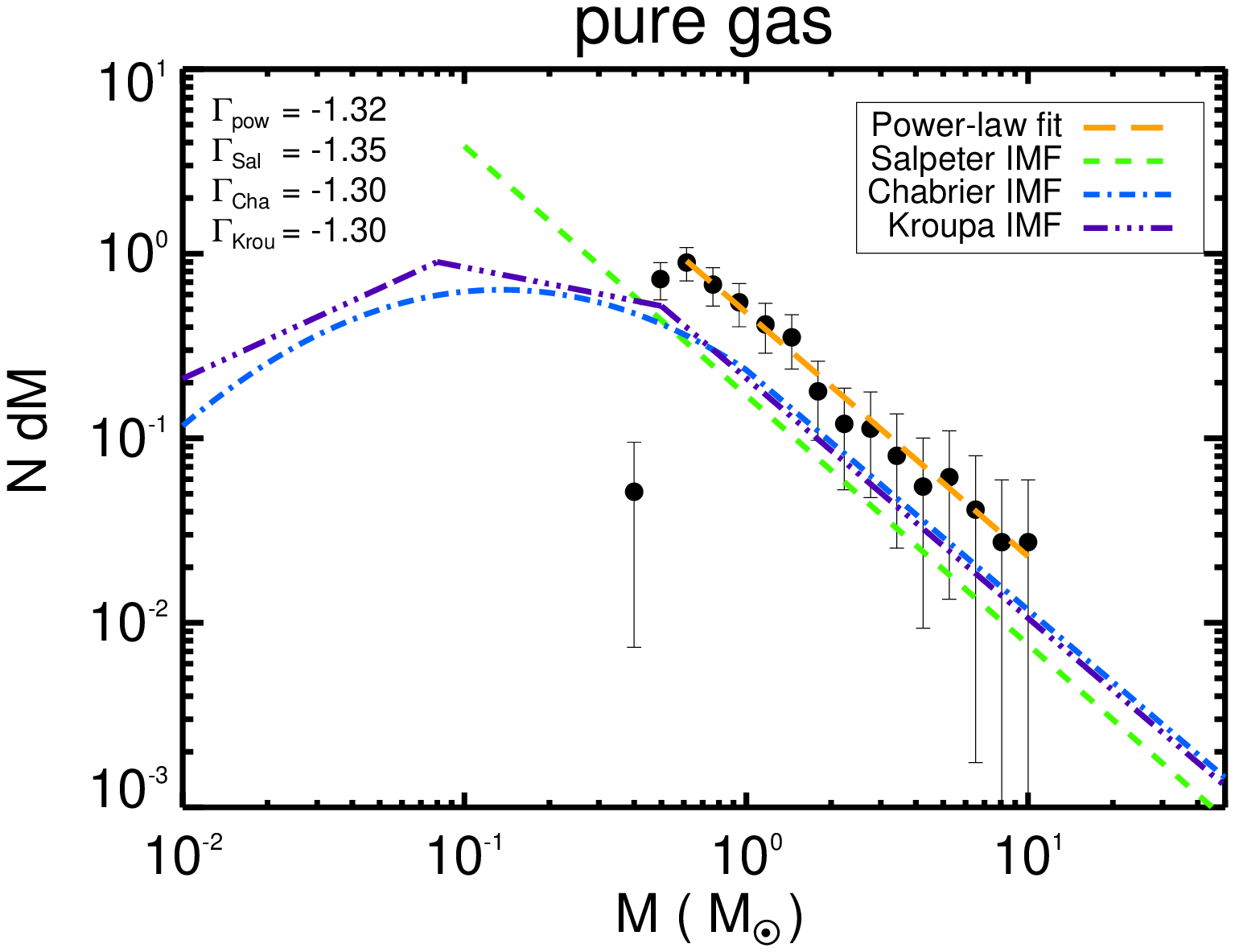}
\includegraphics[scale=0.31]{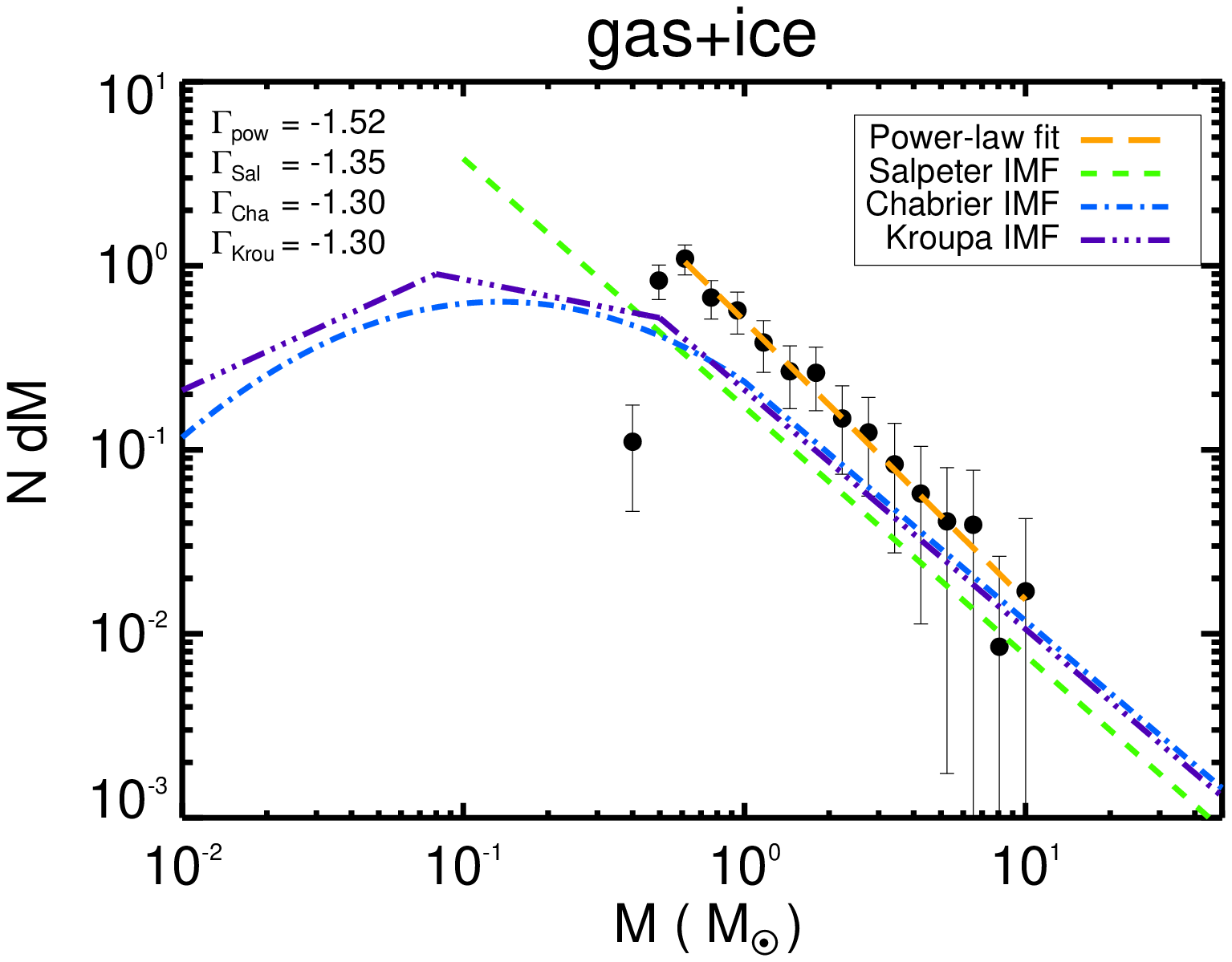}
\caption{The effect of ices on the IMF. A fit through the data (black dots) gives the high-mass slope of the IMF (orange dashed line). Adapted from Hocuk \etal\ (\cite{hocuk3}).}
\label{fig4}
\end{figure*}

Despite a rather significant impact on the thermodynamics, the effect on the initial mass function (IMF) turns out to be minimal. Star-formation seems to be less prone to the chemical influences on cloud dynamics. In Hocuk \etal\ (\cite{hocuk3}), we find a 7\% difference in the amount of stars formed in our simulations in the two extreme cases, a pure gas phase model and a model with freeze-out where desorption is not allowed. Figures \ref{fig3} and \ref{fig4} show the star-formation efficiencies (SFEs), which seem to be largely unaffected, and the IMFs for the two extreme cases. Only a slight deviation in the slope of the IMF is found by the increase in low-mass stars due to the softer EOS. When also non-thermal desorption channels are considered, such as photodesorption and chemical desorption, the difference is even less pronounced.
%
%



\begin{thebibliography}{99}
\bibitem[2014]{2014Natur.505..186F} Fischer, D. B. \etal\ 2014, Nat, 505, 186
\bibitem[1968]{habing1968} Habing, H.J., 1968, bain, 19, 421-+
\bibitem[2014]{hocuk1} Hocuk, S., Cazaux, S., Spaans, M., 2014, MNRAS, 438, 56
\bibitem[2015]{hocuk2} Hocuk, S., Cazaux, S., 2015, A\&A, 576, 49
\bibitem[2015]{hocuk3} Hocuk, S., Cazaux, S., Spaans, M., Caselli, P., 2015, MNRAS submitted
\end{thebibliography}
\end{document}